\documentclass[11pt]{article}
\usepackage{moriond,epsfig}

\input{babarsym.tex}
\RequirePackage{xspace}

\newcommand{\calB}{\ensuremath{{\cal B}}}

\newcommand\etal{{\it et al.}}

\newcommand{\bfig}{\begin{figure}[htbpc!]}
\newcommand{\efig}{\end{figure}}
\newcommand\bef{\begin{figure}}
\newcommand\edf{\end{figure}}

\newcommand\beq{\begin{equation}}
\newcommand\eeq{\end{equation}}
\newcommand\bear{\begin{array}}
\newcommand\enar{\end{array}}
\newcommand\beqa{\begin{eqnarray}}
\newcommand\eeqa{\end{eqnarray}}
\newcommand\ben{\begin{enumerate}}
\newcommand\een{\end{enumerate}}

\newcommand{\fetapiz}{\ensuremath{\eta\piz}\xspace}
\newcommand{\etapiz}{\ensuremath{\Bz\ra\fetapiz}\xspace}
\newcommand{\Betapiz}{\ensuremath{\calB(\etapiz)}\xspace}

\newcommand{\Retapiz}{\ensuremath{(\retapiz)\times 10^{-6}}\xspace}
\newcommand{\uletapiz}{\ensuremath{xx}\xspace}

\newcommand{\fetapKz}{\ensuremath{\etapr K^0}}
\newcommand{\etapKz}{\ensuremath{\Bz\ra\fetapKz}}

\newcommand{\fetappiz}{\ensuremath{\etapr\piz}\xspace}
\newcommand{\etappiz}{\ensuremath{\Bz\ra\fetappiz}\xspace}
\newcommand{\Betappiz}{\ensuremath{\calB(\Bz\ra\etapr \piz)}\xspace}

\newcommand{\Retappiz}{\ensuremath{(\retappiz)\times 10^{-6}}\xspace}
\newcommand{\uletappiz}{\ensuremath{xx}\xspace}

\newcommand{\fetapeta}{\ensuremath{\etapr\eta}}
\newcommand{\etapeta}{\ensuremath{\Bz\ra\fetapeta}}
\newcommand{\Betapeta}{\ensuremath{\calB(\etapeta)}}

\newcommand{\Retapeta}{\ensuremath{(\retapeta)\times 10^{-6}}}
\newcommand{\uletapeta}{\ensuremath{xx}}

\def\Journal#1#2#3#4{{#1} {\bf #2}, #3 (#4)}

\def\PLB{{\em Phys. Lett.}  B}
\def\PRL{\em Phys. Rev. Lett.}
\def\PRGTP{\em Prog. Theor. Phys.}
\def\PRD{{\em Phys. Rev.} D}

\def\ra{\rightarrow}

\def\be{\begin{equation}}
\def\ee{\end{equation}}
\def\bea{\begin{eqnarray}}
\def\eea{\end{eqnarray}}

\def\stwobg   {\ensuremath{\sin(2\beta+\gamma)}\xspace}

\def\KKstarzb {\ensuremath{\Kbar^{(*)0}}\xspace}

\def\DDstarz {\ensuremath{D^{(*)0}}\xspace}

  \renewcommand{\Retapiz}{\ensuremath{0.6^{+0.5}_{-0.4}\pm 0.1}}	
  \renewcommand{\uletapiz}{\ensuremath{1.3}}			

  \renewcommand{\Retappiz}{\ensuremath{0.8^{+0.8}_{-0.6}\pm 0.1}}	
  \renewcommand{\uletappiz}{\ensuremath{2.1}}			

  \renewcommand{\Retapeta}{\ensuremath{0.2^{+0.7}_{-0.5}\pm 0.4}}	
  \renewcommand{\uletapeta}{\ensuremath{1.7}}			
\def\dzkzBrVal  {5.3}
\def\dzkzBrStat {0.7}
\def\dzkzBrSyst {0.3}
\def\dstarzkzBrVal  {3.6}
\def\dstarzkzBrStat {1.2}
\def\dstarzkzBrSyst {0.3}
\def\dzbkstarzBrVal  {4.0}
\def\dzbkstarzBrStat {0.7}
\def\dzbkstarzBrSyst {0.3}
\def\dzkstarzLim    {1.1}
\def\brscale{\ensuremath{\times10^{-5}}}

\def\Btilde {\ensuremath{\tilde{B}}}
\def\Bztilde {\ensuremath{\Btilde^{0}}}
\def\Ktilde {\ensuremath{\tilde{K}}}

\def\Kztilde  {\ensuremath{\Ktilde^{0}}\xspace}

\begin{document}
\vspace*{4cm}
\title{NEW DEVELOPMENTS IN MEASUREMENTS OF CP VIOLATION}

\author{ G. BENELLI }

\address{Department of Physics, The Ohio State University, 191 W. Woodruff Avenue,\\
Columbus OH-43210-1117, USA}

\maketitle\abstracts{
We present several alternative techniques used by the \babar\ 
Collaboration in order to measure the Unitarity Triangle angle 
$\gamma$. We also present the results of two searches designed 
to improve the measurements of $\sin(2\beta)$ using penguin \B\ 
decay modes by reducing the hadronic corrections uncertainties.}

\section{Introduction}
With the discovery of \CP\ violation in the decays of neutral $B$
mesons~\cite{s2b-discovery} and the precise
measurement~\cite{sin2beta} of the angle $\beta$ of the
Cabibbo-Kobayashi-Maskawa~(CKM) Unitarity Triangle~\cite{CKM}, the
experimental focus at the $B$ factories has shifted
towards over-constraining the unitarity triangle through precise 
measurements of \Vub\ and the angles $\alpha$ and $\gamma$, and 
towards measurements of \CP\ asymmetry in the charmless modes, 
that are sensitive to contributions from new physics. The angle 
$\gamma$ is ${\rm arg}(-V_{ub}^{*}V_{ud}/V_{cb}^{*}V_{cd})$ where 
$V_{ij}$ are CKM matrix elements.
Several methods have been suggested and used to measure 
$\gamma$~\cite{gamma-status}, but they all require large samples of 
$B$ mesons not yet available in order to reduce uncertainties. Here 
we present the latest alternative methods explored by the $\babar$ 
experiment to measure the Unitarity Triangle (UT) angle $\gamma$.

The measurement of $\sin(2\beta)$ using penguin modes, while expected 
to yield values consistent with the most precise charmonium measurements, 
showed a small discrepancy in the recent past. In order to 
estimate if this deviation is due to new physics effects, the hadronic 
corrections uncertainties (Standard Model pollution) need to be 
reduced. We present an analysis aimed at reducing such uncertainties 
and the analysis of a new channel, free of Standard Model (SM) pollution.

\section{Alternative methods to measure the Unitarity Triangle angle $\gamma$}

A comprehensive review of $\gamma$ measurements with the standard methods 
currently used was presented by A. Bondar (see his contribution 
to these Proceedings). In this section we will report the alternative 
methods explored by the $\babar$ collaboration and the latest results. 
In general, all measurements of the angle $\gamma$ exploit the quantum 
interference between $b{\to}u$ and $b{\to}c$ transitions that starting 
from the same initial state lead to a common final state. The $b{\to}u$ 
transition, that is sensitive to the weak phase $\gamma$, is 
Cabibbo suppressed with respect to the $b{\to}c$ transition. As a 
consequence the ratio of the amplitudes of these transitions is 
generally small, leading to a difficult measurement. While for charged 
$B$ decays~\cite{bmtodkm,bztodkpi} the time independent asymmetry 
measurement is sensitive to the weak phase $\gamma$ directly, in the case 
of neutral $B$ decays~\cite{sin2bgtheo}, 
since $\BzBzb$ mixing is involved, the relevant weak phase is $2\beta+\gamma$. 
We will first consider the time dependent analyses of neutral $B$ 
decays, then we will look into the time independent analyses of charged 
$B$ decays. 

\subsection{Time dependent CP asymmetries in $\Bz{\to} D^{(*)\pm}\pi^{\mp}$ 
and $\Bz{\to} D^{\pm}\rho^{\mp}$ decays}

\begin{figure}[t]
\begin{center}
\begin{minipage}[h]{7.3cm}
\epsfysize=2.1cm
\epsfbox{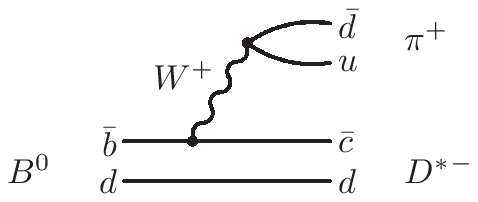}
\end{minipage}
\hskip 1.2cm
\begin{minipage}[h]{7.3cm}
\epsfysize=2.1cm
\epsfbox{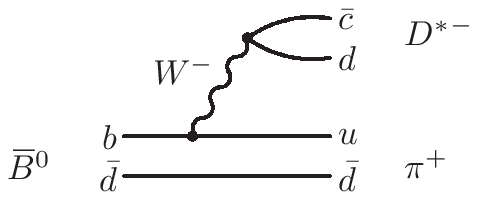}
\end{minipage}
\caption{Feynman diagrams for the Cabibbo favored decay 
$\Bz{\to} D^{(*)-}\pi^{+}$ (left), corresponding to the decay amplitude 
$A_{c}$, and the Cabibbo suppressed decay $\Bzb{\to} D^{(*)-}\pi^{+}$ 
(right), whose amplitude is $A_{u}$.
\label{fig:radish}}
\end{center}
\end{figure}

The time dependent $CP$ asymmetry in the decay modes 
$\Bz{\to} D^{(*)\pm}\pi^{\mp}$ has been extensively studied by 
$\babar$~\cite{sin2bgfull,sin2bgpartial} and BELLE~\cite{belles2bg}. 
In these modes the $CP$ asymmetry arises from the phase difference 
between two amplitudes (see for example Fig.~\ref{fig:radish}): one, 
involving $\BzBzb$ mixing, that is Cabibbo suppressed by the product 
of two small CKM elements ($V_{ub}$ and $V_{cd}$), the other one is 
favored (proportional to $V_{cb}$). Since the parameter $r_{B}^{(*)}$ 
for these decays is close to 0.02, the resulting interference between 
the two amplitudes, proportional to $r_{B}\sin{2\beta+\gamma\pm\delta}$ 
is very small. 
The data sample currently available is not large enough to determine 
the parameter $r_{B}^{(*)}$, but assuming the validity of $SU(3)$ symmetry, 
factorization and excluding $W$ exchange diagrams, it could be 
extracted from the measurement of the branching fraction 
$\calB(\Bz{\to} D_{s}^{(*)\pm}\pi^{\mp})$~\cite{sin2bg_r_rrho}. For the first time 
in this analysis the new decay mode $\Bz{\to} D^{\pm}\rho^{\mp}$, with 
the same diagram, has been studied. With a time dependent maximum 
likelihood fit the following 
results~\cite{sin2bgfulllast} are obtained for the parameters related to the $CP$ violation angle 
$2\beta+\gamma$:
$a^{D\pi}=-0.010 \pm 0.023  \pm  0.007$,
$c_{\rm lep}^{D\pi}=-0.033 \pm 0.042 \pm 0.012$,
$a^{D^*\pi}=-0.040 \pm 0.023 \pm  0.010$,
$c_{\rm lep}^{D^*\pi}=\phantom{-} 0.049 \pm 0.042   \pm  0.015$,
$a^{D\rho}=-0.024 \pm 0.031 \pm  0.009$,
$c_{\rm lep}^{D\rho}=-0.098\pm 0.055 \pm  0.018$.
Combining these results with the partial reconstruction $\Bz{\to} D^{*\pm}\pi^{\mp}$ ones~\cite{sin2bgpartial} the following limits are extracted: $|\stwobg|\!>\!0.64$ $(0.40)$ at $68\%$ $(90\%)$ C.L.~\cite{sin2bgfulllast}

\subsection{Exploring $B^0\to D^{(*)+} a_{0(2)}^-$ decays}

Since the main difficulty with the $\stwobg$ measurement is the small 
value of $r_{B}^{(*)}$, new decay modes have been proposed~\cite{Dsa0_theo} 
to use other two-body final states. The basic idea is that decay 
amplitudes with light scalar or tensor mesons, such as $a_{0}^{+}$ or 
$a_{2}^{+}$, emitted from weak currents, are suppressed due to the 
small coupling constants $f_{a_{0(2)}}$. This means that the Cabibbo 
favored process ($b\to c$ transition) is expected to be suppressed, 
and can be comparable with the Cabibbo suppressed one, resulting in 
a potentially large $CP$ asymmetry. 
\begin{figure}[h]
\begin{center}
\begin{minipage}[h]{6.2cm}
\epsfysize=2.5cm
\epsfbox{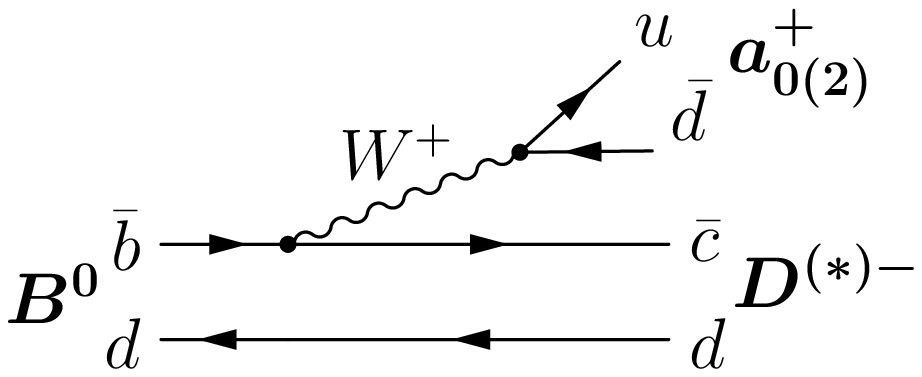}
\end{minipage}
\hskip 0.5cm
\begin{minipage}[h]{6.2cm}
\epsfysize=2.5cm
\epsfbox{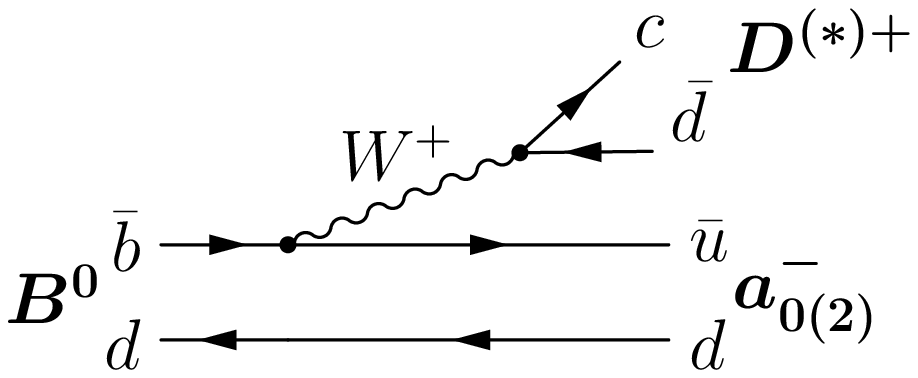}
\end{minipage}
\end{center}
\caption{Tree diagrams contributing to the decay amplitude of 
$B^0 \ra D^{(*)-} a^+_{0(2)}$}
\label{fig:intro}
\end{figure}
Assuming $SU(3)$ symmetry and factorization one can relate
these decays to $\Bz\to D_{s}^{(*)+} a_{0(2)}^-$ decays, where 
the $D$ is substituted by a $D_{s}$, and the branching fractions 
of these are Cabibbo enhanced. A search for $\Bz\to D_{s}^{(*)+} 
a_{0(2)}^-$ has been performed~\cite{Dsa0} 
using about 230 million $\FourS$ decays into $\BzBzb$.
No evidence of these decays has been observed and $90\%$ C.L. 
upper limits on the branching fractions have been set: 
${\cal B}(B^0\to D_s^+ a_0^-) < 1.9\times 10^{-5}$, 
${\cal B}(B^0\to D_s^{*+} a_0^-) < 3.6\times 10^{-5}$, 
${\cal B}(B^0\to D_s^+ a_2^-) < 1.9 \times 10^{-4}$, and 
${\cal B}(B^0\to D_s^{*+} a_2^-) < 2.0\times 10^{-4} $.
The upper limit value for $B^0\to D_s^{+} a_0^-$ is lower than the 
theoretical expectation, which might indicate the need to revisit 
the $B \ra a_0 X$ transition form factor estimate. It might also 
imply the limited applicability of the factorization approach for 
this decay mode. The upper limits suggest that the branching 
ratios of $B^0 \ra D^{(*)+} a_{0(2)}^-$ are too small for 
$CP$ asymmetry measurements given the present
statistics of the $B$ factories.

\subsection{Another idea: $\Bzb\ra \DDstarz \KKstarzb$ decays}

\begin{figure}[htb]
\begin{center}
\epsfig{file=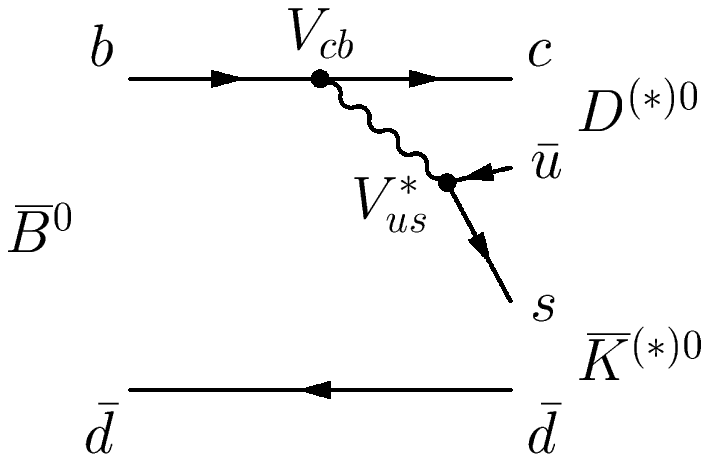,width=0.25\linewidth}
\hskip 1.5cm
\epsfig{file=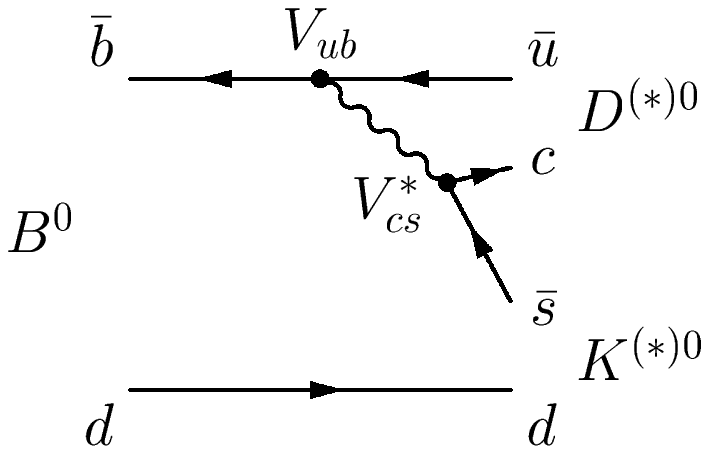,width=0.25\linewidth}
\end{center}
\caption{
The decay diagrams for the $b\ra c$ transition $\Bzb\ra\DDstarz\Kzb$
and the $\bbar \ra \ubar$ transition $\Bz\ra\DDstarz\Kz$.
}
\label{fig:feyn}
\end{figure}

The decay modes $\Bzb\ra\DDstarz\Kzb$ offer a new approach for the
determination of \stwobg\ from the measurement of time-dependent
\CP\ asymmetries in these decays~\cite{bmtodkm}.
The \CP\ asymmetry appears as a  result of the interference
between two diagrams leading to the same final state
$\DDstarz\KS$~(Figure~\ref{fig:feyn}).
A $\Bzb$ meson can either decay via a $b\ra c$ quark transition to
the $\DDstarz\Kzb$ ($\Kzb\to\KS$) final state, or oscillate into a
$\Bz$ which then decays via a $\bar b\ra \bar u$ transition to the
$\DDstarz\Kz$ ($\Kz\to\KS$) final state (reference to the charge 
conjugate state is implied).
The $\Bzb\Bz$ oscillation provides the weak phase $2\beta$ and
the relative  weak phase between the two decay diagrams is $\gamma$.

\begin{figure}
\begin{center}
\epsfig{file=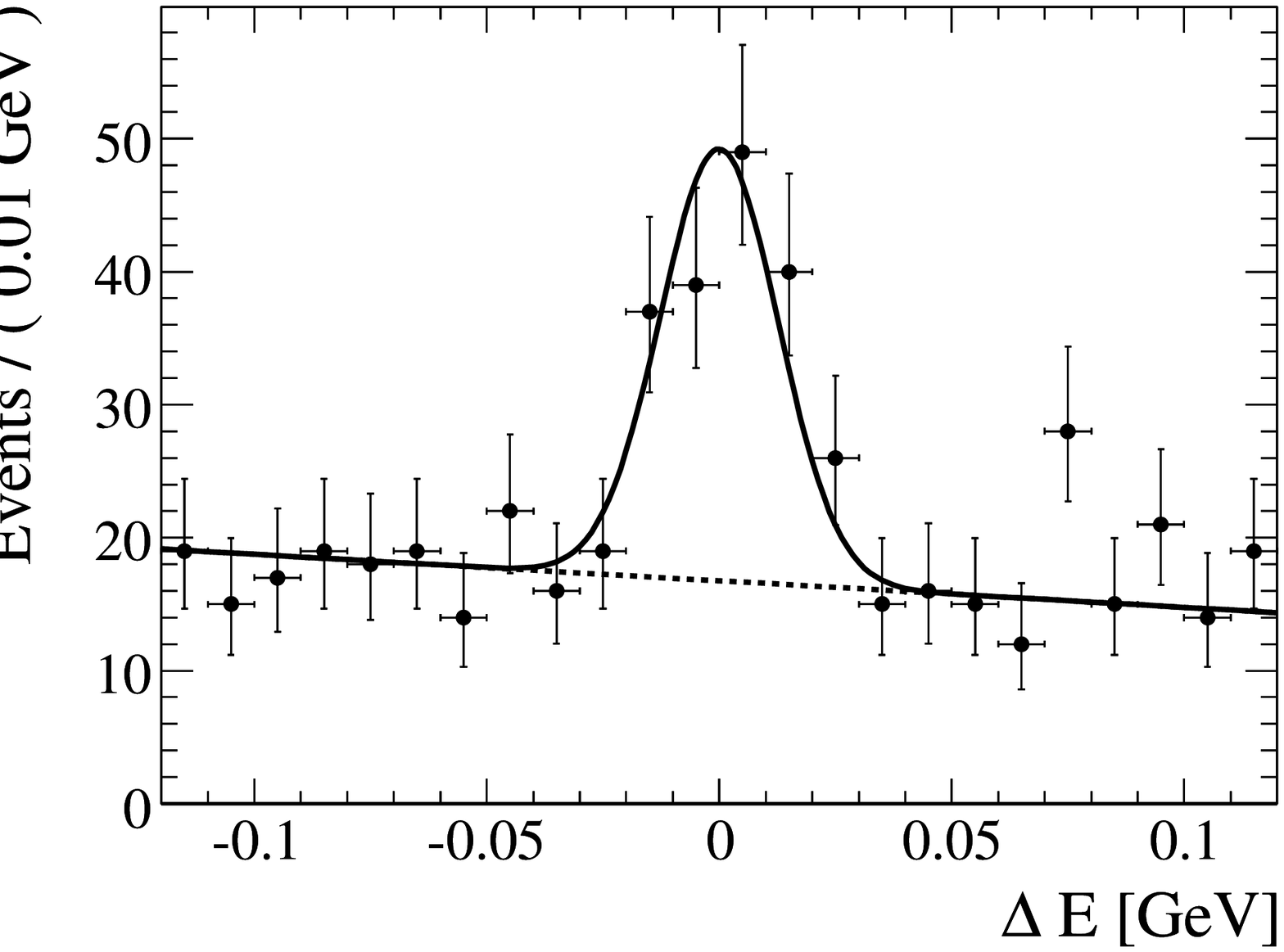,width=0.31\linewidth}
\put(-102,74){a)}
\hskip 1.0cm
\epsfig{file=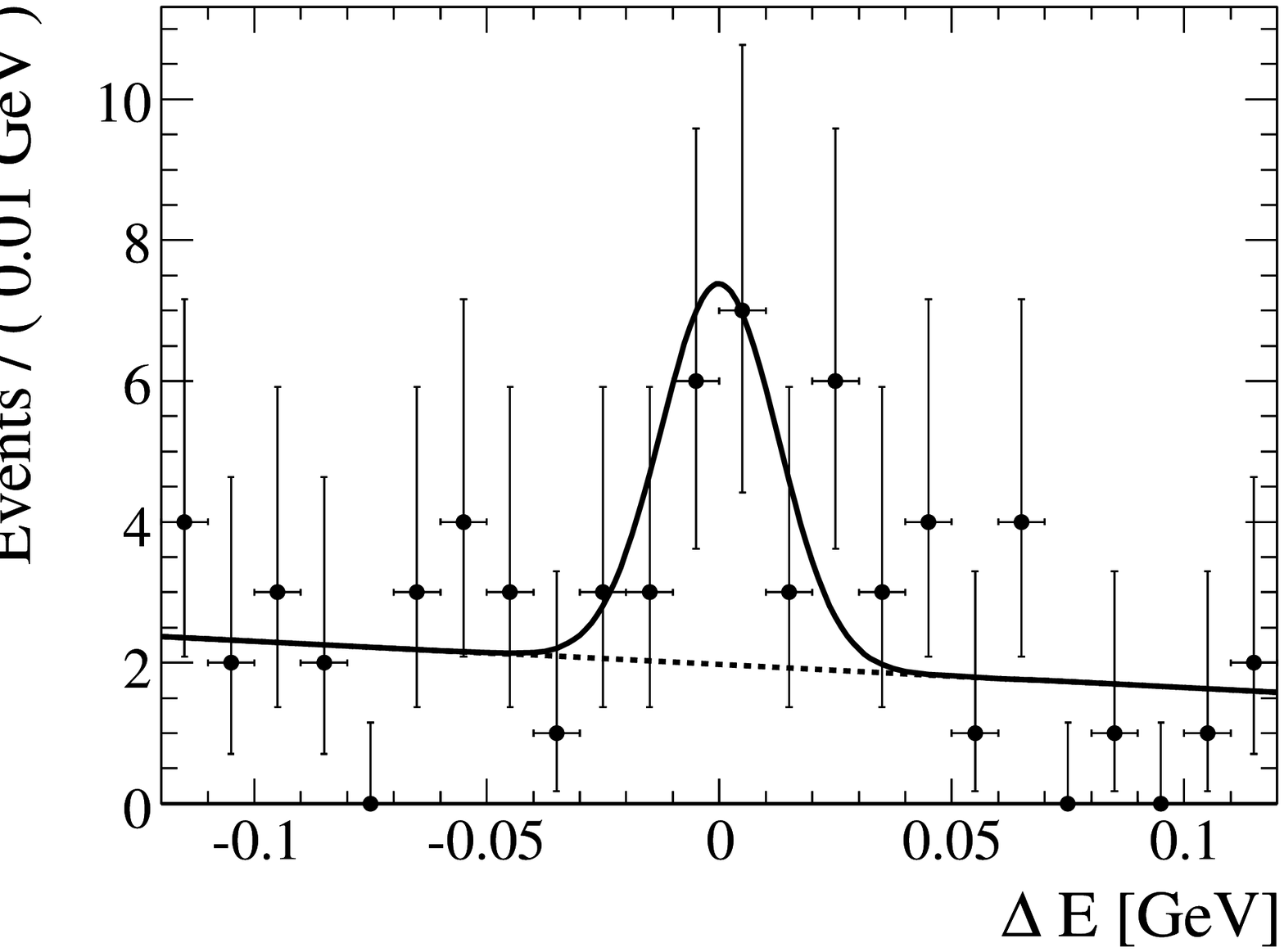,width=0.31\linewidth}
\put(-102,74){b)}\\
\epsfig{file=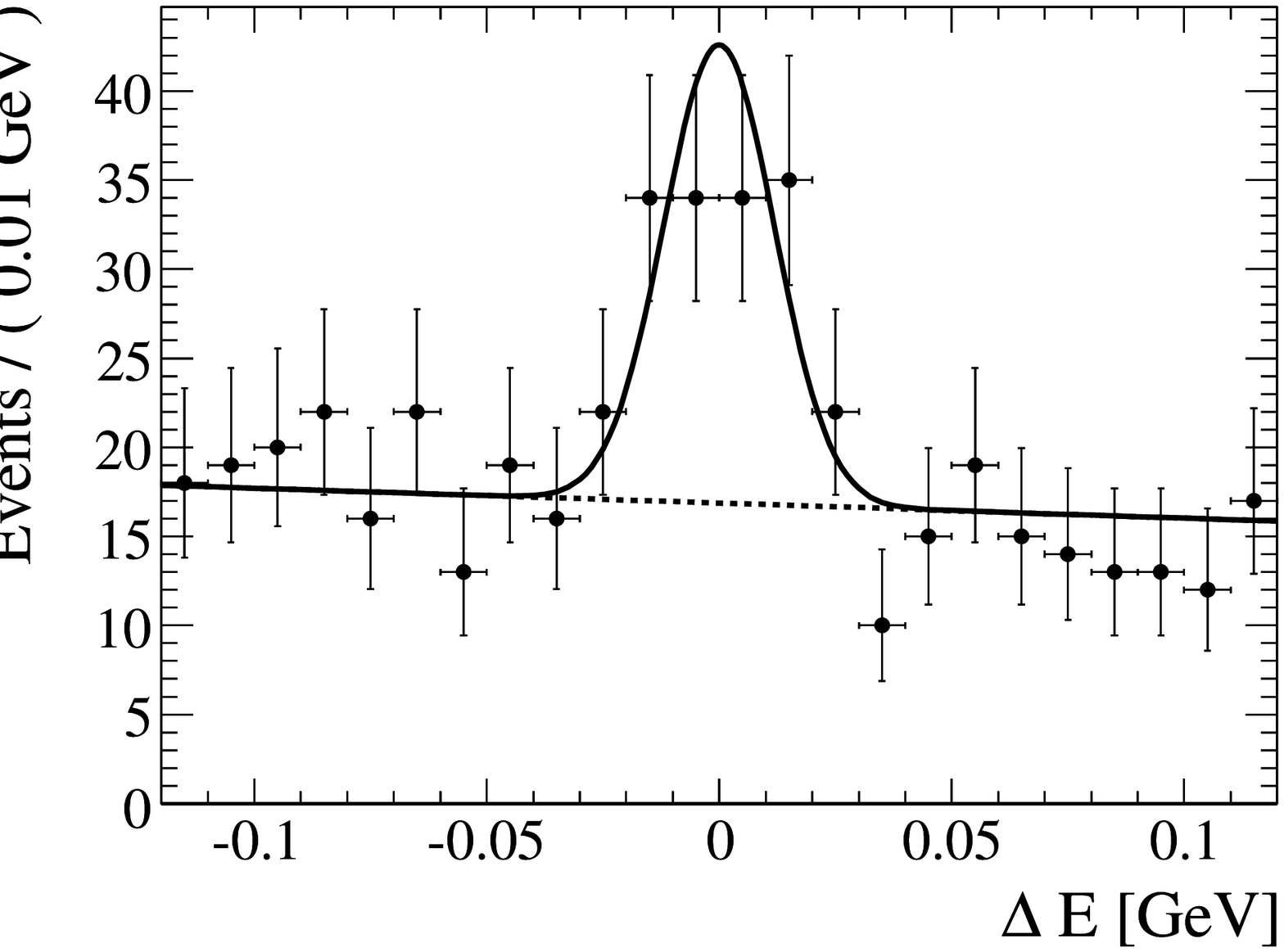,width=0.31\linewidth}
\put(-102,74){c)}
\hskip 1.0cm
\epsfig{file=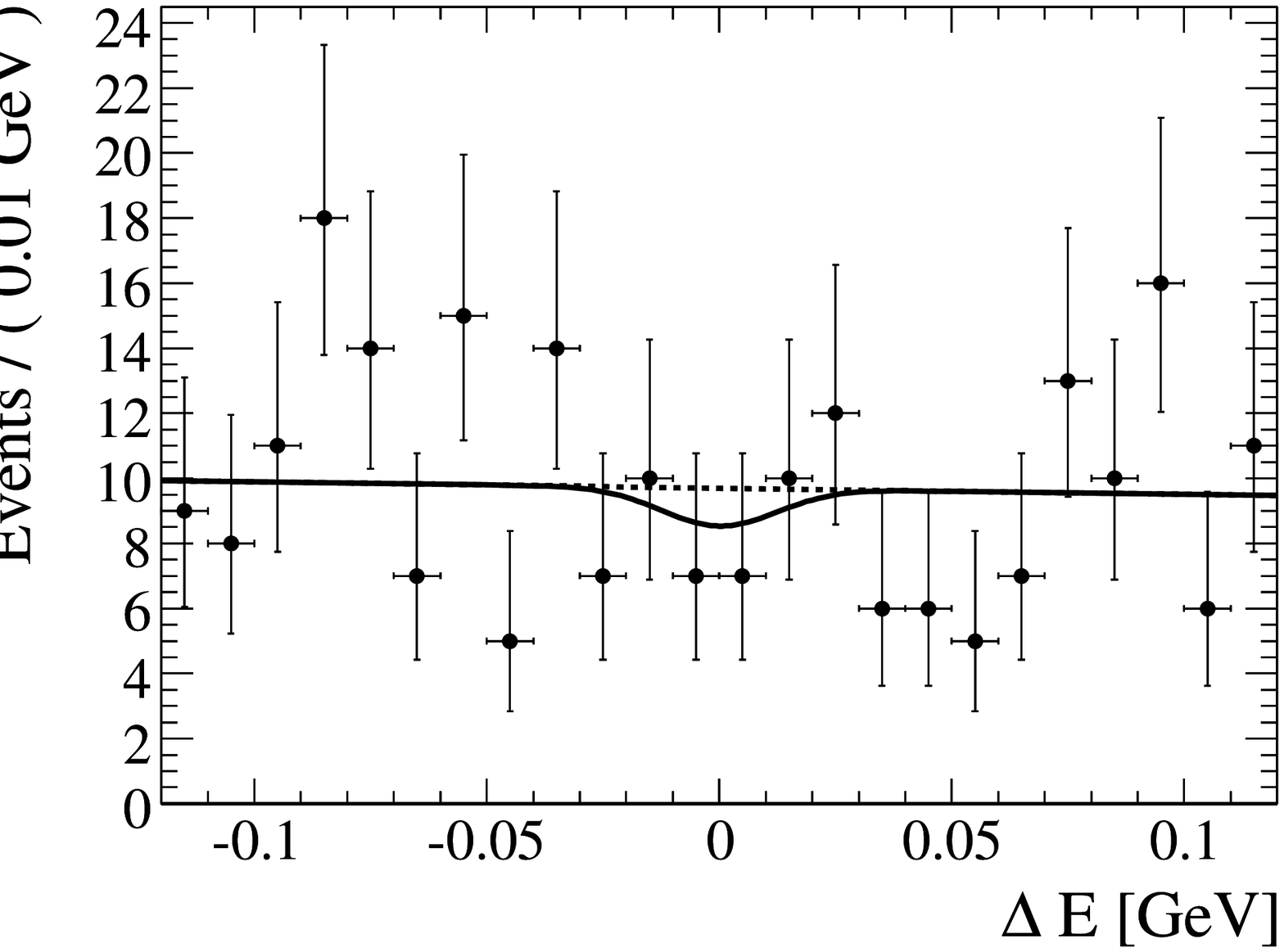,width=0.31\linewidth}
\put(-102,74){d)}
\caption{
Distribution of \DeltaE\ for  a) $\Bztilde\ra\Dz\Kztilde$, b)
$\Bztilde\ra\Dstarz\Kztilde$, c) $\Bzb\ra\Dz\Kstarzb$, and d)
$\Bzb\ra\Dzb\Kstarzb$ candidates with $|\mes-5280~\mevcc|< 8\mevcc$. The
points are the data, 
the solid curve is the projection of the likelihood fit, and
the dashed curve represents the background component.
}
\label{fig:yields}
\end{center}
\end{figure}

Both diagrams are color suppressed, this means that the parameter $r_{B}$ 
could be large (expected to be close to $0.4$), so the $CP$ asymmetry is 
expected to be large. In order to get insight into the Cabibbo suppressed 
diagram in $\Bz\to D^{(*)0}\Kzb$, the self-tagging $\Bz\to D^{(*)0}\Kstarzb$ 
with $\Kstarzb\ra\Km\pip$. Using a sample of 226 million $\BzBzb$ decays, 
the Cabibbo favored processes have been observed~\cite{D0K0}. The 
\DeltaE\ distributions of candidates for the sums of the reconstructed 
\Dz\ decay modes are illustrated in Figure~\ref{fig:yields} and no evidence 
of the Cabibbo suppressed mode $\Bz\to \Dzb\Kstarzb$ was observed.
The following branching fractions are obtained: 
$\BR(\Bztilde\ra\Dstarz\Kztilde)=(\dstarzkzBrVal\pm\dstarzkzBrStat\pm\dstarzkzBrSyst)\brscale$, 
$\BR(\Bztilde\ra\Dz\Kztilde)=(\dzkzBrVal\pm\dzkzBrStat\pm\dzkzBrSyst)\brscale$,
and 
$\BR(\Bzb\ra\Dz\Kstarzb)=(\dzbkstarzBrVal\pm\dzbkstarzBrStat\pm\dzbkstarzBrSyst)\brscale$. 
A 90\% confidence level upper limit is set on 
$\BR(\Bzb\ra\Dzb\Kstarzb) < \dzkstarzLim\brscale$.
These limit can be translated into an upper limit for the decay amplitude ratio 
$r_{B}\equiv|{\mathcal A}(\Bzb\ra\Dzb\Kstarzb)/{\mathcal A}(\Bzb\ra\Dz\Kstarzb)|$
to be less than 0.4 at the 90\% confidence level, excluding the naive expected
value for $r_{B}$.

\subsection{Time independent $CP$ asymmetries in $\Bz\ra\Dzb\Kp\pim$ and $\Bz\ra\Dz\Kp\pim$ decays}

Even if these decays are from neutral $B$ mesons, this analysis is time 
independent and it is sensitive directly to $\gamma$, not $2\beta+\gamma$.
The principle is the same of all time independent charged $B$ analyses: two 
diagrams (Cabibbo favored $b\to c$ and Cabibbo suppressed $b\to u$ 
transitions) interfering, one of which is color favored, the other 
color suppressed. The new idea behind this analysis is that in three 
body decays such as $\Bz\ra\Dzb\Kp\pim$ and $\Bz\ra\Dz\Kp\pim$, it is 
possible to get two extra $b\to c$ and $b\to u$ transitions, adding a color 
allowed $b\to u$ transition diagram~\cite{dalitz,bztodkpi}. This could lead 
to larger rates and 
potentially significant $CP$ asymmetries. In addition, a Dalitz plot analysis of
the $DK\pi$ final state can resolve the strong phase and reduce the ambiguity
to two-fold, compared to the GLW standard method with the charged $B$.
Using a sample of 226 million $\BB$ events, a branching fraction measurement 
of the Cabibbo favored decay has been performed~\cite{DKPi}: 
${\cal B}(\Bz\ra\Dzb\Kp\pim) = (88 \pm 15 \pm 9 )\times 10^{-6}\;$. Resonant 
contributions were observed but the search for the Cabibbo suppressed 
$\Bz\ra\Dz\Kp\pim$ decays showed no evidence of signal. A $90\%$ C.L. 
upper limit on the branching fraction was set: 
${\cal B}(\Bz\ra\Dz\Kp\pim) < 19 \times 10^{-6}\,$. The event yields in 
these modes are lower than expected, indicating that a larger data sample
is required in order to constrain $\gamma$ with this type of analysis.

\section{New developments in the measurement of \boldmath{$\sin{2\beta}$} using penguin modes\label{sec:penguin}}

Measuring $\sin{2\beta}$ with penguin modes opens the possibility of observing, 
or constraining new physics phenomena that could enter the loops. 
If the decay amplitude of these penguin modes is dominated, as naively 
expected, by the short distance penguin transition $b\to s\sbar s$, 
then the measured asymmetry in the tree level and penguin case should be consistent. 
Unfortunately, for many of these penguin modes there are other Standard Model 
diagrams that introduce the so-called SM pollution. Understanding the 
contribution from these hadronic corrections and reducing their uncertainties 
is thus necessary in order to evaluate any eventual contribution from 
new physics. To address this problem, several measurements have been proposed to 
estimate these corrections, and at the same time penguin modes that are 
theoretically free from SM pollution have been explored. Here we report 
the results of some branching fractions measurements needed to reduce the SM pollution
uncertainties mentioned above, and the analysis of a new pollution-free penguin mode
by the $\babar$ collaboration.

\subsection{Branching Fraction Limits for \Bz\ Decays
to \fetapeta, \fetappiz\ and \fetapiz}

The branching fraction measurement of these $\Bz$ decays into two-body 
charmless final states has its own interest in comparing with the various
theoretical predictions from QCD
factorization \cite{BBNS}, perturbative QCD (for
$\Bz\ra\eta^{(\prime)}\piz$) \cite{pQCD}, soft collinear effective
theory \cite{SCET}, and flavor-SU(3) symmetry 
\cite{CGR}.  
The expectations lie in the approximate ranges $0.2$--$1.0\times10^{-6}$ for
$\Bz\ra\eta^{(\prime)}\piz$, and $0.3$--$2\times10^{-6}$ for \etapeta.
These decays are also of interest in constraining the expected value of
the time-dependent \CP-violation asymmetry parameter $S_f$ in the decay with
$f=\etapr\KS$~\cite{CGR,GLN}.  
The leading-order SM calculation gives the equality
$S_{\etapr\KS} = S_{J/\psi\KS}$, where the latter has been
precisely measured \cite{sin2beta}, and 
equals \stwob\ in the SM. The \CP\ asymmetry in the charmless modes
is sensitive to contributions from new physics, but also to
contamination from sub-leading SM amplitudes.
The most stringent constraint on
such contamination in $S_{{\etapr}\KS}$ comes from the measured
branching fractions of the three decay modes
studied in this paper~\cite{CGR,GLN}.  Recently it has also
been suggested~\cite{GZ} that
\etappiz\ and \etapiz\ can be used to constrain the
contribution from isospin-breaking effects on the value of \stwoa\ in
$B\to\pi^+\pi^-$ decays.
No evidence of any of the signal signatures above was found in the 232 
million $\BB$ pairs collected by $\babar$.
Combining the
measurements we obtain the central values and
90\%\ C.L. upper limits for the branching fractions:
$\Betapeta = (\Retapeta ) \times 10^{-6}\ (<\uletapeta\times 10^{-6})$, 
$\Betapiz = (\Retapiz ) \times 10^{-6}\ (<\uletapiz \times 10^{-6})$, and 
$\Betappiz = (\Retappiz ) \times 10^{-6}\ (<\uletappiz\times 10^{-6})$.
These upper limits represent
two to three-fold improvement over the previous measurements
\cite{Compare}. 
The range of sensitivity of these measurements is comparable to the
range of the theoretical estimates.
Using the method proposed by Gronau \etal~\cite{GLN}, these results 
will provide approximately $20\%$ improvement of the prediction for the 
contribution of the color suppressed tree amplitude in \etapKz\ decays.  
This translates into a 20\% reduction of this theoretical uncertainty 
in $S_{\etapr\KS}$.  A similar improvement is found in the corresponding 
uncertainty of \stwoa\ measured with $B\to\pi^+\pi^-$ decays~\cite{GZ}. 

\subsection{Exploring $\Bz\to\KS\KS\KL$ penguin decays}

As mentioned above, recent CP asymmetry measurements in 
$b\to s\qbar q$ penguin decays have suggested deviations from the SM 
expectations. The final state is a $CP$ 
eigenstate, so there is no $CP$ dilution effects, and it is also a 
pure $b\to s\sbar s$ penguin transition, so it is free from SM pollution~\cite{KsKsKL_theo}. 
This channel has been searched for the first time~\cite{KsKsKL} and no 
evidence was found for this 
decay in 232 million $\BB$, so a branching fraction measurement and upper
limit have been derived: $\calB(\KS\KS\KL)=(2.4^{+2.7}_{-2.5}\pm 0.6)\times10^{-6}$,
$\calB(\KS\KS\KL)<6.4\times10^{-6}$ at 90$\%$ C.L.

\section{Conclusion}

The measurement of $\gamma$ at the B factories is currently still limited by 
statistical uncertainties and a larger sample of $B$ mesons is needed to improve the
measurement. In the meanwhile a lot of effort has been spent in finding alternative
paths to $\gamma$, all of which proved to lack statistical significance with the current
data sample.
Improvements in the estimate of hadronic corrections uncertainties has been made 
that will allow a more clear interpretation of some of the $\sin{2\beta}$ penguin results.
A new SM pollution free analysis has been performed but the channels proved to be suppressed beyond the initial theoretical predictions.

\end{document}